\RequirePackage{fix-cm}
\documentclass[smallextended]{svjour3}       
\smartqed  
\usepackage{graphicx}
\usepackage{soul} 
\usepackage[usenames,dvipsnames]{xcolor}
\usepackage{rotating} 

\newcommand{\lya}{Ly$\alpha$~}

\journalname{Experimental Astronomy}
\begin{document}

\title{On the feasibility of studying the exospheres of  Earth-like exoplanets by Lyman-$\alpha$ monitoring.\thanks{This work has been funded my the Ministry of Economy and Competitivity of Spain under grant numbers ESP2014-54243-R and ESP2015-68908-R}
}
\subtitle{ Detectability constraints for nearby M stars.}

\titlerunning{On the detectability of Earth-like exoplanets in Ly$\alpha$}        

\author{Ana I G\'omez de Castro         \and
             Leire Beitia-Antero \and
             Sabina Ustamujic
}

\authorrunning{G\'omez de Castro et al. } 

\institute{ Ana I. G\'omez de Castro \at AEGORA Research Group, Universidad Complutense de Madrid,
              Plaza de Ciencias 3, 28040 Madrid, Spain\\
              \email{aig@ucm.es}           
           \and
           Leire Beitia-Antero \at  AEGORA Research Group, Universidad Complutense de Madrid,
              Plaza de Ciencias 3, 28040 Madrid, Spain\\
            \and
           Sabina Ustamujic \at  AEGORA Research Group, Universidad Complutense de Madrid,
           Plaza de Ciencias 3, 28040 Madrid, Spain\\ 
}

\date{Received: date / Accepted: date}

\maketitle
\begin{abstract}

Observations of the Earth's exosphere have unveiled an extended envelope of hydrogen reaching
further than 10 Earth radii  composed of atoms orbiting around the Earth.  This large envelope
increases significantly the opacity of the Earth to Lyman $\alpha$  (\lya) photons coming from the Sun,
to the point of making feasible the detection of the Earth's transit signature from 1.35 pc  if pointing with an 8~meter primary 
mirror space telescope through a clean line of sight ($N_H < 10^{17}$~cm$^{-2}$), as we show. 
In this work, we evaluate the  potential detectability of Earth analogues
orbiting around nearby M-type stars by monitoring the variability of the \lya flux. We show
that, in spite of the interstellar, heliospheric and astrospheric absorption, the transit signature in M5 V type stars would
be detectable  with a dedicated \lya flux monitor implemented in a  4-8 m class space telescope.  Such monitoring 
programs would enable measuring the robustness of planetary atmospheres under heavy space weather conditions 
like those produced by M-type stars. A 2-m class telescope, such as the World Space 
Observatory, would suffice to detect an Earth-like planet orbiting around Proxima Centauri,  if there
was such a planet or nearby M5 type stars. 

\end{abstract}




\section{Introduction} \label{sec:intro}

Atomic hydrogen is the dominant constituent of
the upper terrestrial atmosphere. The
primary source of atomic hydrogen is photodissociation by
solar ultraviolet photons of molecular species (H$_2$O, CH$_4$, H$_2$)
originated in the troposphere that are transported to higher
altitudes (Brasseur \& Solomon 1996, Dessler et al. 1994).  
Above 90 km, in the thermosphere, direct
production of atomic hydrogen is balanced by upward
diffusion with an estimated total escape flux of $10^8$~cm$^{-2}$~ s$^{-1}$.
This permanent loss of cold hydrogen atoms has a significant
impact on long-term atmospheric evolution (Shizgal \& Arkos, 1996). \par

Recent measurements have shown evidence of hot
hydrogen atoms in the upper layers of the thermosphere.
This population is generated by non-thermal processes and traces the
coupling of  the upper atmospheric
layers to the plasmasphere, the magnetosphere and
to space weather conditions in general (Qin \& Waldrop, 2016).  The
interaction of hydrogen atoms with hot protons from
the Solar wind and the energetic ions trapped in the
ion belts lead to charge-exchange reactions and the
production of energetic hydrogen atoms (1- 10$^3$ keV) that
are key actors in the ion-neutral coupling between the
atmosphere and magnetosphere. Hence, the precise
determination of the atmospheric hydrogen
distribution is vital for investigations of the
chemistry and the thermal and particle flow in the
Earth's atmosphere, as well as its coupling with Solar activity. \par

The distribution of exospheric hydrogen is studied through the
Lyman~$\alpha$ (Ly$\alpha$) transition, the most sensitive tracer to  thin
columns of neutral gas.  Space probes have detected the Ly$\alpha$ photons 
from the Sun scattered by the hydrogen atoms in the Earth's exosphere. 
Missions such as the Magnetopause-to-Aurora Global 
Exploration (IMAGE) satellite (Fuselier et al. 2000) or the Two Wide-angle 
Imaging Neutral-atom Spectrometers (TWINS) (McComas et al. 2009) have 
carried out these measurements while navigating  within  the exosphere.
From these observations we know that there exists an extended component
of the exosphere, significantly stronger than predicted by
the classical model (Chamberlain 1963).
Extended exospheres have also been detected  in Venus and Mars  
(Shizgal \& Arkos, 1996) and it seems to be a common characteristic of terrestrial planets. \par	

The detection of exoplanets has opened the possibility of studying
planetary exospheres in other planetary systems  submitted to 
diverse stellar radiation fields and space weather conditions.
The transit of planetary exospheres in front of the stellar disk
produces a net absorption that has been detected in the 
stellar Ly$\alpha$  profile.  Since the seminal work by
Vidal-Madjar et al. (2003) who detected the signature of 
HD 209458b exosphere,   Ly$\alpha$ absorption 
has also been detected from the hot Jupiter HD 189733b (Lecavelier Des Etangs et al. 2010)
and from the warm Neptune GJ 436b (Kulow et al. 2014, Ehrenreich et al. 2015). 
However, it has not been detected in the two super-Earths
observed to date , namely  55 Cnc  e (Ehrenreich 2012) and HD 97658b (Bourrier et al. 2017).
They have hypothesised  that the lower extreme ultraviolet radiation from these stars 
diminishes the hydrogen
production rate in the thermosphere and hence, the extent of the
exosphere. In fact, it is possible that extended, Earth-like
exospheres are not common or even detectable with the current techniques.

Measuring the impact of dramatic enhancement of the solar activity on planetary exospheres
had been unthinkable until the detection of Earth-like exoplanets. 
These are orbiting around  M dwarfs
with strong magnetic fields and winds, thereby providing the ideal scenario
to test Solar System-based planetary wind models under strong conditions. 
In this work, we evaluate the Earth's transmittance  to solar Ly$\alpha$
photons  and compute  theoretical Ly$\alpha$ light curves for the transit of  
an Earth-like planet\footnote{An 'Earth-like' planet is assumed to have the 
same extended exosphere as the Earth's.} orbiting around an M-dwarf star. We show that a 4-8~m 
class space telescope should be able to detect these variations provided the 
transit signature is not absorbed by the Interstellar Medium (ISM). 
In Section 2 the transmittance of the Earth exosphere to Ly$\alpha$ photons is evaluated.
In Section 3 the light curves are computed for Earth-like planets orbiting some
sample stars. The potential of Ly$\alpha$ monitoring for the study of planetary exospheres
is discussed in Section~4. A brief summary is provided in Section~5.

\section{Transmittance of the Earth's exosphere to Ly$\alpha$ photons}

The GEO instrument (Mende et al. 2000) on board the IMAGE satellite
found that the hydrogen density distribution is essentially
cylindrically symmetric around the Sun-Earth line, exhibiting an
enhancement in the antisolar direction, towards the
geotail ({{\O}}stgaard et al. 2003, hereafter O2003). 
Along any given solar zenital angle, the density
distribution is bimodal with a dominant central (or core) component
and an extended component peaking at $\sim 8 R_{E}$. \par

Later on, the TWINS mission obtained the 3D distribution of H~I atoms between 3R$_{E}$ 
and 8R$_{E}$ and reported enhancements by a factor of $\sim 2-3$ of the 
extended component with respect to O2003  (Zoennchen
et al. 2010).

From these works, the distribution of hydrogen in the Earth's exosphere can
be modelled   as,

\begin{equation}
  n_{H}(r) = n_0 e^{- \frac{r}{r_0}}+\kappa n_1 e^{-\frac{r}{r_1}} 
\end{equation}

\noindent
for any given solar-zenital angle.  $r$ is the radial distance to the center of the Earth,
$(n_0,r_0)$ are the density and characteristic radius of the core
component, dominated by the  Earth photoevaporative flow, and $(n_1,r_1)$ 
are the density and characteristic radius of the extended component.
$\kappa$ is a coefficient introduced to modify at will the relative strength of the 
extended component with respect to the core component. Minimum values of these 
coefficients are: $n_0 = 10^4$~cm$^{-3}$,  $n_1 = 7$~cm$^{-3}$, 
$r_0 = 1.02 R_E$, $r_1 = 8.2 R_E$ and $\kappa = 1$ (O2003). 

Using  this distribution as a baseline, the column density of exospheric hydrogen 
($N_{H}$) and the transmittance of the Earth's exosphere to stellar Ly$\alpha$ 
photons for a distant observer located on the plane of the ecliptic can be directly 
worked out.  At a given projected distance, $R$, from the center of the Earth, 
the stellar  Ly$\alpha$ photons propagate through a layer of thickness 
$2z_l = 2 (r_e^2-R^2)^{1/2}$ , with $r_e$ the radius  of the exosphere taken 
to be 15 $R_{E}$ for this calculation.  The output radiation, $I(R)$, can be calculated 
from the standard radiation transfer equation,

\begin{equation}
  \frac{dI}{I} = - \sigma n(r) dz,
\end{equation}
\noindent
being $z = \sqrt{r^{2}-R^{2}}$ the depth of gas layer
through which the Ly$\alpha$ radiation is being transferred and $\sigma = 5.9\times10^{-12}\times 1050^{-1/2}$~cm$^2$ (Bishop, 1999) the cross
section of hydrogen to Ly$\alpha$ radiation; 1050~K is the reference temperature of the
exospheric hydrogen (Bishop 1999, O2003).  

This equation can be numerically integrated  and for $dz<<\sigma n(r)$ takes the form,

\begin{equation}
  4\pi I(R) = \int_{-z_{l}}^{+z_{l}}I_{0}\sigma n(r) dz
\end{equation}
\noindent

The transmittance of the Earth's exosphere at a given projected distance is defined as,

\begin{equation}
Tr(R) = \frac{I(R)}{I_0}
\end{equation}
\noindent
and it  is plotted in Fig.~1 for various possible $n(r)$ distributions; the reference 
curve is based on IMAGE/GEO results (O2003). We observe that at distances $>3-4 R_{E}$, the
transmittance is very sensitive to the strength of the extended
component. Additional transmittance curves enhancing the strength
of the extended component by a factor of $\kappa=5$ to 40  with
respect to O2003 are also represented; $\kappa=5$ mimics well the
results from TWINS (Zoennchen et al. 2010). 

\begin{figure}[h!]
  \centering
  \includegraphics[width = \linewidth]{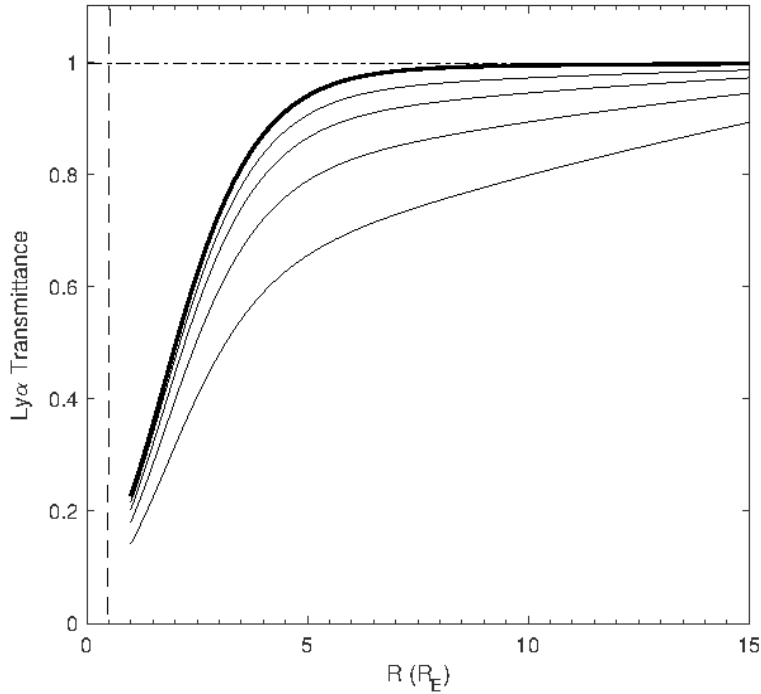}
  \caption{Transmittance of the Earth's exosphere to
    Ly$\alpha$ photons in normal incidence from the Sun as a
    function of the radial distance to the center of the
    Earth (units: Earth radii ($R_{E}$)). $\kappa$ indicates the
    relative strength of the extended component with
    respect to the baseline model obtained from IMAGE/GEO
    data (O2003). $\kappa = 1$ corresponds to the baseline model (bold).
    Incremental $\kappa$ values: 5, 10, 20, 40 are represented with thin lines. 
    Measurements based on TWINS suggest $\kappa=5$ 
    (Zoennchen et al. 2010). The  geometric cross-section of the 
    solid Earth to solar Ly$\alpha$ photons is represented for guidance
    (dashed line). 
}
  \label{fig:1}
\end{figure}

\subsection{The impact of the ISM on the observability}

During transit observations, the exospheric \lya  absorption is expected to be maximum at the core of the stellar 
\lya  profile. Unfortunately, the radial velocity of the 
nearby M type stars is similar to that of the local clouds in the ISM
which absorb efficiently the core of the line (see {\it e.g.} Youngblood et al. 2016).
As a result, though large Earth-like exospheres  may produce a measurable signature during transit
in the \lya  light curves (see below), this would pass unnoticed if the core of the line is
fully absorbed by the ISM; a hydrogen column of $N_H \simeq 5 \times 10^{17}$~cm$^{-2}$ suffices 
to block the \lya  flux (see {\it e.g.} G\'omez de Castro et al. 2016).  \par

 The lines of sight with measured \lya absorption are plotted in Figure 2 from the compilations by Wood et al. 2005 and
Redfield and Linsky 2008. The location of the Local Interstellar Cloud (LIC) is neatly traced. Additional clouds and filaments are 
also identified. The distribution in the sky of the  local \lya absorbers is compared with the distribution of M stars in the 
northern hemisphere as compiled for the CARMENES input catalogue\footnote{ CARMENES is conducting a 600-night 
exoplanet survey targeting ~300 M dwarfs during Guaranteed Time Observations. The main scientific objective of CARMENES is to carry out a survey of late-type main sequence stars with the goal of detecting low-mass planets in their habitable zones. In the focus of the project are very cool dwarf stars later than spectral type M4 and moderately active stars.} (Alonso-Floriano et al. 2015).  Figure~2 also shows that the information on the characteristics and spatial distribution of the local \lya absorbers is rather scarce.

\begin{figure}[h!]
  \centering
  \includegraphics[width = \linewidth]{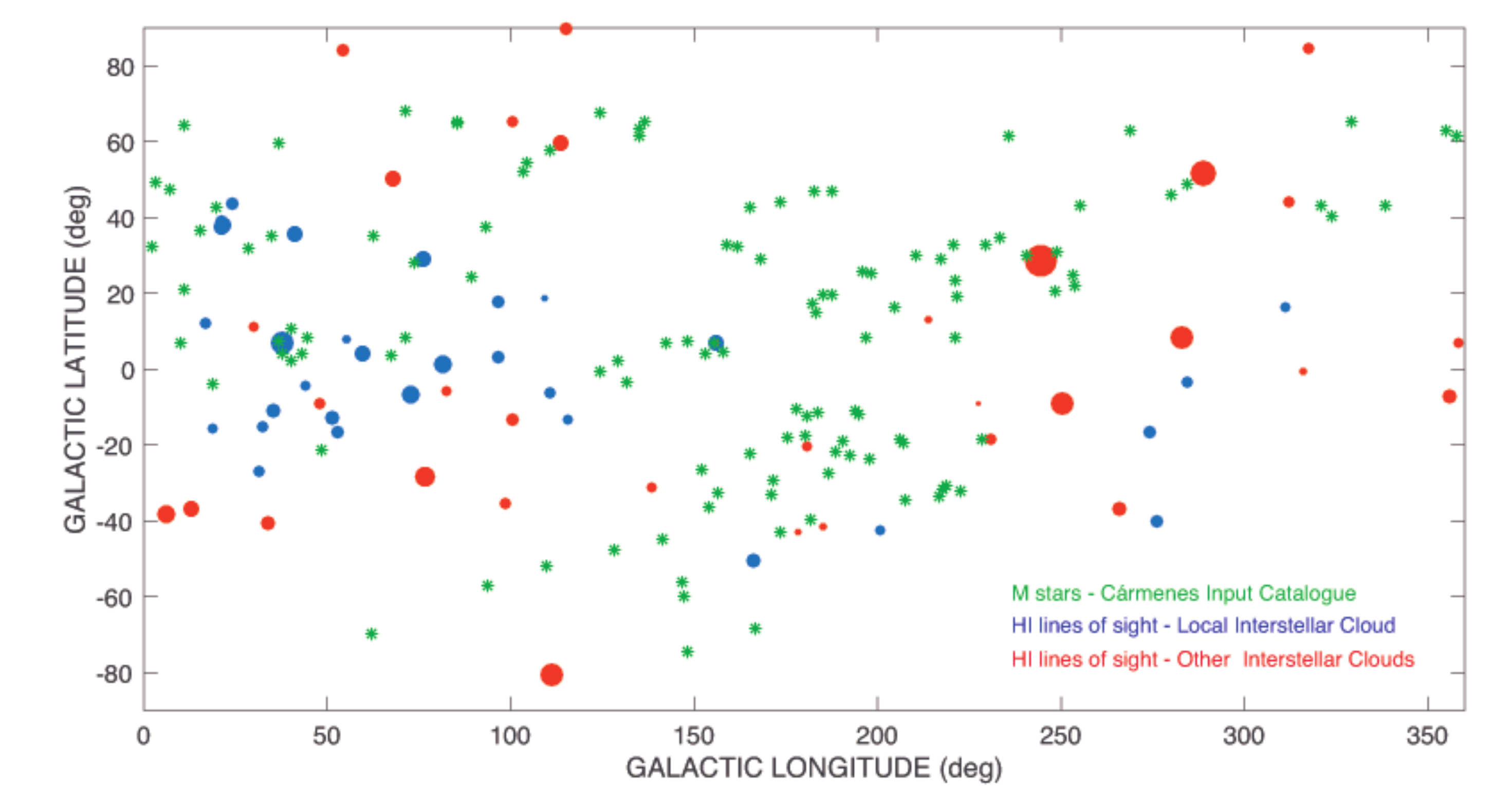}
  \caption{Distribution in the sky of the \lya absorbers as per works by Wood et al. 2005 and Redfield and Linsky 2008 compared with 
  the location of the M-type stars in the CARMENES input catalogue (marked with star symbols). Absorbers are marked with dots. The size of dots scales with the
  HI column density ($N_H$), from $N_H = 10^{17.2}$~cm$^{-2}$ for the smallest dots to $N_H = 10^{18.6}$~cm$^{-2}$ for the largest
  ones. 
}
  \label{fig:2}
\end{figure}

The existence of diffuse clouds in the ISM  however, has not prevented the detection of hot Jupiters transits (see \textit{e.g.} 
Vidal-Madjar et al. 2003) because exospheres are heated by the stellar XUV radiation fields\footnote{XUV stands for X-ray and
extreme UV photons.} and disturbed by the interaction with the stellar wind. The XUV flux is efficiently
absorbed by the upper atmospheric layers and heats the gas favouring the thermal escape. As a result, 
a significant fraction of the exospheric hydrogen may reach velocities high enough to allow the detection in spite of the 
ISM absorption of the line core. The high optical depth of the \lya  line also assists the detection of the 
feature since the lowermost exospheric layers saturate the  core of the \lya profile and produce extended wings. \par

To evaluate the impact of high stellar XUV fluxes on the dynamics of Earth-like planets' exospheres 
we have resourced to the models developed by Erkaev et al. (2013); these authors analysed the atmospheric escape 
by solving the equations of a hydrodynamical flow including the volume heating rates by stellar 
soft X-ray and UV radiation. Outputs from the model are the density ($n(r)$), 
temperature ($T(r)$) and expansion velocity ($V_e(r)$) from the thermosphere to the outer exosphere. 
Models are calculated for a broad range of XUV fluxes, spanning from the level of the quiet Sun (XUV$_{\odot}$) to 
100 times this value\footnote{The solar XUV flux at 1 AU is XUV$_{\odot}$= 4.64 erg cm$^{-2}$ s$^{-1}$ (Ribas et al., 2005).}.
For instance, the XUV fluxes irradiating the exoplanets GJ 436b and  GJ 674b are 355~erg cm$^{-2}$ s$^{-1}$ (76 XUV$_{\odot}$) and 
3631~erg cm$^{-2}$ s$^{-1}$ (783 XUV$_{\odot}$) respectively  (Sanz-Forcada et al. 2011). Both GJ 436 and GJ 674 
are M2.5 V stars. Recently, France et al. (2016) have derived  an XUV flux of 10 - 70 cm$^{-2}$ s$^{-1}$ in the habitable 
zone of typical exoplanet host stars. \par

Using as input the functions $n(r)$, $T(r)$ and $V_e(r)$  from Erkaev et al. 2013, we have computed 
the optical depth of the exospheric absorption of \lya  photons 
as a function of $R$, the cylindrical distance to the center of the Earth-like planet and the velocity 
of the exospheric gas $v$, as,  

\begin{equation}
\tau (R,v) = \sigma \int_{-z_{l}}^{+z_{l}} n(r) f(r,v) dz
\end{equation}

\noindent
with $f(v,z)$ the normalized velocity profile of the gas at a distance $r$ obtained by the convolution
of the thermal broadening (assuming T(r) as in Erkaev et al. 2013)
and the natural broadening profile ($\phi (v)$). Hence,
 
\begin{equation}
f(r,v)= \frac {\exp(-U(r,v))} {2k\pi T(r)/m_H} \ast \phi(v)
\end{equation}

\noindent
and,
\begin{equation}
U(r,v) = \frac {(v-V_e(r))^2}{2k T(r) /m_H}
\end{equation}

\noindent
Note that $m_H$ is the mass of a hydrogen atom, $k$ is the Boltzmann constant and
\begin{equation}
r= \sqrt{(R^2+z^2)}
\end{equation}

\noindent
The transmittance is computed as per Eq. 4. As shown in  Figure 3, the exospheric transmittance
displays a strengthening of the blue shifted wing as both the stellar XUV flux and the escape flow increase.
Also, the high optical depth of the \lya line  produces broad wings that result on a 33\% of the transmittance occurring 
at Doppler shifts larger than $\pm$40~km~s$^{-1}$ for an XUV flux equal to XUV$_{\odot}$
 (this value raises to a 58\% for a XUV flux = 50 XUV$_{\odot}$).  Therefore, the 
 detectability is strongly dependent on the XUV flux.  Stable exospheres are more difficult to detect than unstable, 
 photoevaporating exospheres forced by the action of a strong stellar XUV radiation field. Note that the high XUV radiation
 from M-type stars may result in the full evaporation of the  planetary exosphere/atmosphere rendering fruitless any atmospheric
 research on these sources. Otherwise, the investigation of the atmospheres stability under heavy XUV irradiation is a fundamental
for  planetary/exoplanetary research.

\begin{figure}[h!]
  \centering
  \includegraphics[width=\linewidth]{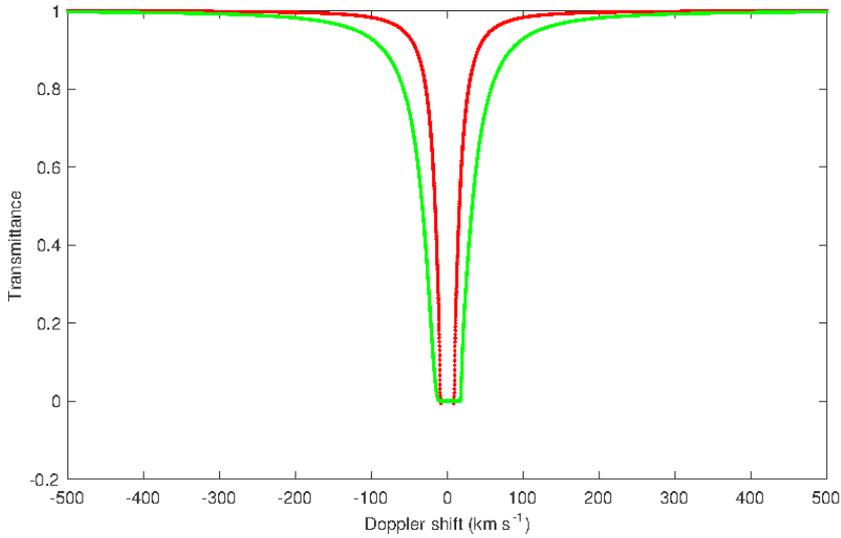}
  \caption{Transmittance of the atmosphere of an Earth-like planet to stellar \lya photons as a function of the
  Doppler shift with respect to \lya rest wavelength. Atmospheric models come from Erkaev et al. 2013.
  The red line represents the transmittance of an exosphere irradiated with an XUV flux alike the
  produced by the quiet Sun. The green line also represents the transmittance but assuming an XUV flux
  50 times higher.}
  \label{fig:3}
  \end{figure}

Detectability also depends on the properties of the individual sources, \lya  profile, ISM absorption,
and radial velocity as illustrated in Fig. 4,  where synthetic profiles are represented for the transit of an 
Earth-like exoplanet orbiting in the habitable zone around M5 and M0 V stars. For the sake of the calculation,
we have selected two nearby  M stars with well measured
Ly$\alpha$ fluxes, namely AU~Mic (M0 V) and Proxima Centauri (M5.5 V) (see Table~1).   In all cases, an 
Earth-like planet is assumed to be orbiting in the habitable zone. 
The stellar \lya profiles have been taken from Wood et al. (2001, 2005) who reconstructed the stellar 
\lya emission after removing the effect of the ISM absorption for a large set of cool stars. In particular,
the AU Mic profile and the Proxima Cen profile have selected to represent M0 and M5 stars. 
In a first step, the absorption produced by the exosphere of the transiting planet is computed 
as per the procedure defined in Eq. 3-8.  In a second step, this output profile is assumed to cross
an interstellar cloud with column density,  $N_H = 10^{18}$~cm$^{-2}$, temperature,  $T= 8,000$~K
and radial velocity with respect to the star, $\Delta V = 0, 15$ and $30$~km~s${-1}$ to evaluate the
impact of the relative velocity between the cloud and the star in the detectability. 
As expected, the transit signal increases as the stellar radius decreases and becomes comparable to the
size of the Earth exosphere.

\begin{table}[h!]
\caption{Parameters used for the transit calculation.}
\begin{tabular}{lll}
\hline\hline
Parameter &   AU Mic & Prox Cen\\ 
\hline
Spectral Type$^{(a)}$&  M0V & M5.5V\\
Radius (R$_{\odot}$)&  0.5$^{(b)}$ & 0.14$^{(c)}$ \\
Mass (M$_{\odot}$) &  0.6$^{(d)}$ & 0.12$^{(e)}$ \\
Semimajor axis of the orbit (a.u.) & 0.3 & 0.032 \\
F$^0_{Ly\alpha}(10^{-12}$erg~cm$^{-2}$s$^{-1}$)$^{(a)}$&10.3& 4.21 \\
F$_{Ly\alpha}(10^{-12}$erg~cm$^{-2}$s$^{-1}$)$^{(a)}$& 2& 1.2 \\
logN$_H$ (cm$^{-2}$)$^{(a)}$ &  17.6 & 18.36 \\
d (pc)$^{(a)}$ &  9.9 & 1.30 \\ \hline
\end{tabular}
\begin{small}
\begin{tabular}{l}
F$^0_{Ly\alpha}$ is the stellar \lya flux reconstructed by Wood et al. 2005.\\
F$_{Ly\alpha}$ is the observed \lya flux calculated from Hubble archival data.\\
References:
(a) Wood et al. 2005; (b) Rhee et al. 2007; \\
(c) Maldonado et al. 2015; (d) Chen et al. 2005;\\
(e) Benedict et al. 2016.\\

\end{tabular}
\end{small}
\end{table}

\begin{figure}[]
  \centering
  \includegraphics[width=\linewidth]{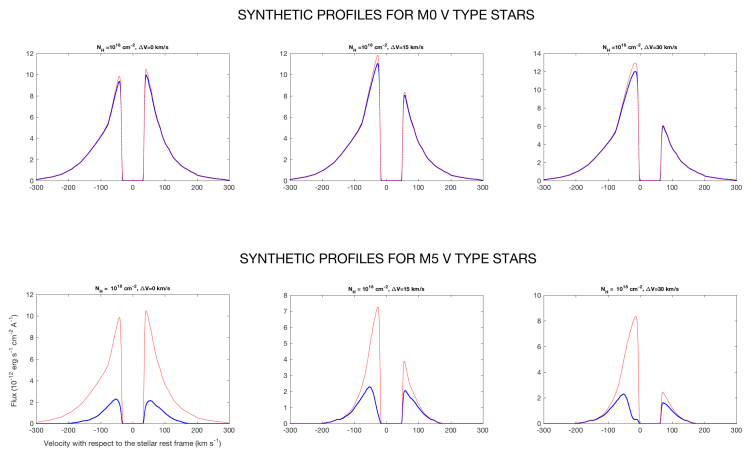}
  \caption{Simulated \lya  profiles for M0 and M5 spectral type stars. The \lya profile
  in the absence of a transiting planet in plotted in red. The blue-solid line represents the profile during transit (at maximum). In all cases,
  it is assumed that there is an ISM cloud in the line of sight with  $N_H = 10^{18}$~cm$^{-2}$ and $T= 8,000$~K. The relative 
  velocity between the cloud and the star is taken to be 0, 15 and 30 km~s$^{-1}$ in the left, central and right panels.}
  \label{fig:4}
  \end{figure}

In summary, the detectability of  \lya  transits of Earth-like exoplanets orbiting M-type stars will depend strongly 
on the stellar spectral type (stellar radius),  the XUV flux that drives the atmospheric outflows and the chance of 
observing along clean lines of sight.  Planets orbiting around high velocity stars, such as 
the Kaptein star, will also offer a good chance for exospheric studies since most of the local ISM clouds have
radial velocities about 10-40 km/s.

\section{Calculation of the transit light curves for some sample stars}

Earth-like planets are being detected by the transit method:
orbiting planets occult the stellar radiation when
crossing the line of sight, producing a measurable variation
of the radiation that reaches the observer. The best suited
stars for this purpose  are M-dwarfs. 
They are the smallest and the most common class of stars in the Galaxy.
GJ 1132b (Berta-Thompson et al. 2015), Kepler-42c (Borucki et al. 2011), GJ 3470b (Bonfils et al. 2012) or
Kepler-1646b (Morton et al. 2016)
are some examples of positive detections of exoplanets around M stars. A candidate to Earth-like planet has been
detected in Proxima Centauri,  at 1.3 pc from the Sun (Anglada-Escud\'e, 2016).
Earth-like exoplanets have also been detected around earlier types of cool stars
such as Kepler-78b (Pepe et al. 2013), a K-type dwarf, or the late G-type star
Kepler-93b (Dressing \& Charbonneau 2015). \par

We have computed  transit light curves for AU~Mic and Proxima~Cen using the parameters in Table~1. 
In addition, we have computed the light curve for $\alpha$-Cen, the nearest 
analogue to the Sun, to evaluate the detectability of the Earth exosphere from a distance of 1.3 pc. The Sun and $\alpha$-Cen 
have similar \lya fluxes and the \lya absorption along the line of sight is well quantified (Wood et al. 2005). 
From the data available in the Hubble archive, we have measured the observed \lya flux of $\alpha$-Cen 
to be $4.2\times10^{-11}$~erg~cm$^{-2}$~s$^{-1}$.  Note that the \lya flux varies within the activity cycle. During the solar cycle,
the \lya irradiance varies\footnote{ https://lasp.colorado.edu/lisirg/lya} roughly between 3.5 and  $6.5\times10^{11}$photons~cm$^{-2}$~s$^{-1}$  
(Rairden et al. 1986). The flux used for our calculations
is similar to that of the active Sun observed from $\alpha$-Cen.

For all the calculations, the Ly$\alpha$ emissivity has been taken as constant over the stellar disk;
the presence of dark spots and active regions has been neglected
though the light curves could also be used to detect them.  In all cases, 
the inclination is assumed to be 90$^o$ (edge-on).  

The light curves have been calculated under the assumption that  impact of the ISM
blockage of the transit signature  is equivalent to degrading the exosphere transmittance
to a 10\% of the original value. According to Fig.~4 this is a reasonable approach for
atmospheres illuminated by high XUV fluxes. The theoretical light curves are shown  Fig. 5.
The light curve varies from source to source due to differences in the stellar radius, transit duration 
and location of the habitable zone (see Table 1).   \par

\begin{figure}[]
  \centering
  \includegraphics[width=5cm]{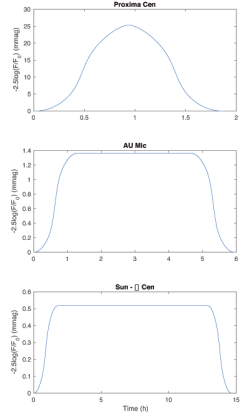}
  \caption{Theoretical  \lya light curves from the transit model for Proxima Cen, AU Mic and $\alpha$ Cen
  assuming that  90\% of the transit signal is absorbed by the ISM (see text).  }
  \label{fig:5}
  \end{figure}

\subsection{Detectability}

To evaluate the feasibility of detecting the light curves in Fig.~5, an
instrumental set-up needs to defined.  We have assumed a rather simple configuration
consisting of a \lya monitor operating in a 4 m primary mirror space telescope.  
The monitor is a low dispersion slitless spectrograph with resolving power 800
serviced by  a photon counting CsI MCP detector; this configuration is  similar to the 
implemented in the Focal Camera Unit (FCU) of the WSO-UV mission
(see Sachkov et al 2016 for details) . A spectrograph is implemented instead of a \lya filter because 
it is the most  efficient  from the radiometric point of view (see, e.g. G\'omez de Castro et al. 2014).
The light path to the detector includes three reflections (primary, 
secondary and pick-off mirror) each with high reflectivity (80\%)
and the low dispersion grating  with
reflectivity 20\% at 122~nm. 
A 20\% detector  quantum efficiency at 122 nm has been considered (Vallerga et al. 2009).  
For the sake of the simulation, we have assumed a stable heliospheric
background and have neglected the geocoronal emission. 
The heliospheric background is taken to be $5.5\times 10^{-3}$~photons $cm^{-2} s^{-1}$  
(Lallement et al. 1996, Koutroumpa et al. 2017).

Note that the Earth geocoronal emission has been neglected because the observation
is assumed to be carried out from a space telescope in
geosynchronous orbit such as the WSO-UV or above. During the
observation the satellite is assumed to be  far from the magnetotail and pointing in the
anti-Earth direction\footnote{This is the optimal configuration: the space telescope
observes the target when it is at opposition with respect to the Earth. This orientation minimizes the geocoronal 
\lya background since most of the hydrogen atoms are located close to the base of the exosphere. 
Along a given line of sight, the contribution of the Earth geocorona to the \lya flux is given by,
\begin{equation}
F_{Ly\alpha} = \int _{r_0} ^\infty I_{Ly\alpha} dr
\end{equation}
According to  O2003, the geocorona \lya ~ intensity follows the
law:  $I_{Ly\alpha}= 18.1 \exp (-r/1.01R_E)+1.05\exp(-r/8.21R_E)$
with $I_{Ly\alpha}$ in kilo Rayleighs (kR). At low Earth orbit (LEO), $r_0 \simeq 6750$~km, $F_{Ly\alpha}$(LEO) = 6.77 kR while in a  
geosynchronous orbit, $r_0=6.6 R_E$, $F_{Ly\alpha}$(GO) =0.08 kR. Thus the contribution is significantly smaller
and lower than  the heliospheric background  (Lallement et al. 1996, Koutroumpa et al. 2017).}

With these provisions, the SNR is given by,
\begin{equation}
SNR = \frac { \Delta SCR \times T_{\rm exp} }{\sqrt{SCR \times T_{\rm exp} +N_{pix} \times BCR \times T_{\rm exp}}}
\end{equation}
\noindent
with
\begin{itemize}
\item $SCR$, the stellar \lya ~ count rate.
\item $BCR$, the \lya ~ background count rate.
\item $\Delta SCR$ measures the depth of transit: the out of transit SCR minus the SCR at maximum depth during transit. 
\item $N_{pix}$ the number of pixels used to integrate the signal. For this calculation, $N_{pix}=4$ is assumed. 
\end{itemize}

The count rate is derived from the \lya flux, $F_{Ly\alpha}$, from the usual expression:

\begin{equation}
SCR = \left( \frac {D}{ 4 m} \right) ^2 A_{eff} \frac {F_{\rm Ly \alpha} } {h \nu _{\rm Ly\alpha}}  
\end{equation}

\noindent
with $A_{eff}$ the effective area,
\begin{equation}
A_{eff} = \pi \left( \frac{400 {\rm }}{2} \right) ^{2} 0.8^3 0.2^2 = 2.6 \times 10^3 {\rm cm}^2
\end{equation}

\noindent
and $(D/4 m)$ a scale factor to increase the effective area with the size (D) of the primary mirror
of the telescope.

SNRs have been computed  neglecting the ISM blockage of the transit signature because its impact in the light curve depends
on many factors\footnote{The properties of the ISM cloud affect the blockage of the transit signal, in particular,
Hydrogen column density, temperature, relative velocity with respect to the star and turbulence.} that depend on
the particular source under study.
To increase the SNR, the counts have been binned into  time steps of 2, 4 and 10 minutes
for Proxima Centauri, AU~Mic and $\alpha$~Cen, respectively. With these provisions the SNR reached 
is:  11.7, 1.2 and 1.5 for Proxima Centauri, AU~Mic and $\alpha$~Cen, respectively. If the size of the primary mirror is 
increased to 12 m ($D/4m = 3$), the SNR rises to 3.5 for AU~Mic and to 4.4 for $\alpha$~Cen; see the light curves in Fig. 6.

\begin{figure}[]
  \centering
  \includegraphics[width=8cm]{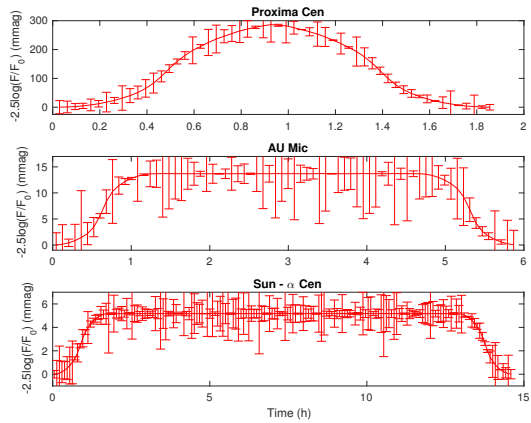}
  \caption{Observed  \lya light curves (no ISM absorption is considered) according to the noise model and observational configuration
  described in Sect.~4. SNR(Proxima Cen) = 11.7 obtained with a 4 m telescope and time binning of the photon counts
  of 2 minutes. SNR(AU Mic) = 3.5  obtained with a 12 m telescope and time binning 4 minutes. SNR($\alpha$ Cen) = 4.4 
  obtained with a 12 m telescope and time binning 10 minutes.   }
  \label{fig:5}
  \end{figure}

  Note that the presence of ISM absorption can be directly accounted in $\Delta SCR$. For instance, the presence of  a \lya absorber
 with  $N_H = 10^{18}$~cm$^{-2}$ and $T= 8,000$~K in the line of sight to an M0 V star decreases 
$\Delta SCR$ by a 96.5-97.5 \% in the range of relative velocities used for the simulated profiles in Fig.~4. 
However, the same absorber in the line of sight to an M5 V star decreases $\Delta SCR$ by barely a factor of 2
(43.8-60.2 \% depending on the cloud velocity). In all cases, the total counts can be raised by co-adding several 
observations of the transit to reach SNR values similar to those quoted above for the non-ISM absorption case. 
This technique is  performed routinely by missions, such as Kepler, devoted to exoplanetary research.

\section{Discussion: Potential exoplanetary studies using a Ly$\alpha$ monitor}

In Sect.~2 and 3.1, we have shown that extended, Earth-like exospheres produce detectable absorption
signature over the stellar  Ly$\alpha$ emission because of the strong opacity of the line and the
large extent of the Earth's exosphere. This signature is especially strong in small stars such as the 
M dwarfs. So far, Ly$\alpha$ observations of exoplanet's exospheres  have only been feasible with the
Hubble Space Telescope (HST) orbiting in LEO. However, a similar size telescope but in geosynchronous (or higher) orbit
would enjoy a much darker (and estable) \lya background allowing to carry out efficient \lya monitoring programs.

WSO-UV is a 170~cm primary mirror space telescope that would be equipped with a solar-blind CsI MCP type detector optimised for the detection of
Ly$\alpha$ photons (Shustov et al. 2015a,b, G\'omez de Castro et al. 2014). The far UV channel of the camera will be equipped with a prism for slitless spectroscopy  providing  a resolving power of 800 in the Ly$\alpha$ line to separate cleanly the line from nearby features. As shown in
Fig.~7, the sensitivity would be high enough to study the extended distribution of the exospheric hydrogen and measure the impact of stellar activity
in Proxima~Cen.  Moreover, these observations will allow studying the planetary magnetic fields. In the Earth, the scale height of the hydrogen distribution increases with declining solar activity suggesting that the upper thermosphere is not a completely thermalized regime (Qin \& Waldrop 2016) and that the energy input from collisions with magnetospheric and solar wind particles cannot be neglected.

\begin{figure}[h!]
  \centering
  \includegraphics[width = \linewidth]{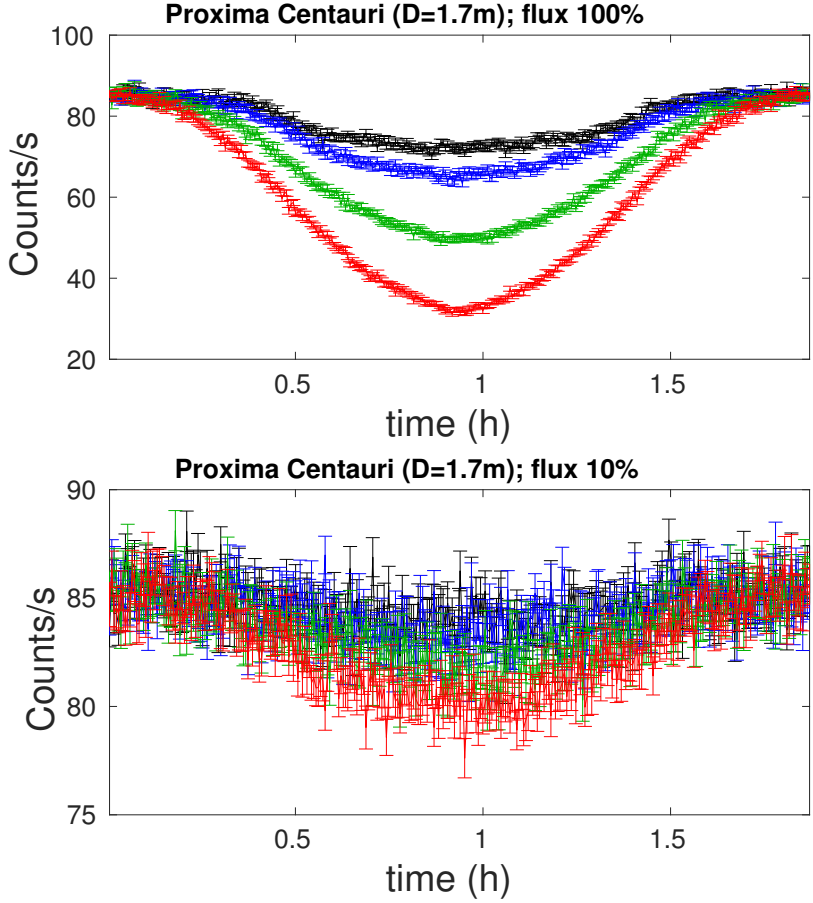}
  \caption{Impact of the distribution of hydrogen
    in the exosphere in the transit light curve of
    Prox Cen as observed by WSO-UV. Models corresponding to
    $\kappa$ = 1 (black), 5 (blue), 10 (green) and 40 (red) are plotted.  The light curve has been calculated for the {\it Case A} and the {\it Case B} assumptions (0\% and 90\% of the \lya flux  is absorbed by the ISM, respectively);  no temporal binning has been applied and 100 transits have been coadded.}
  \label{fig:7}
\end{figure}

However,  WSO-UV is far too small to detect Earth-like planets orbiting in the habitable zone  around stars bigger than a nearby M5V
type star,  no to mention around G2V, solar like stars. Therefore, it is worth extending the calculation for the up-coming generation
of large space telescopes such as the European Ultraviolet Visible Observatory (EUVO) (G\'omez de Castro et al. 2014) or the Large Ultraviolet-Optical-Infrared Observatory (LUVOIR\footnote{https://asd.gsfc.nasa.gov/luvoir/}).

In Figure~8, we have plotted the detectability of transits as a function 
of the telescope size and the stellar Ly$\alpha$ flux for M type stars. The diagram represents  for a given stellar
flux, the combination of exposure time and telescope diameter ($\eta = (D/4 m)^2 \times T_{exp}$) required  to achieve
SNR=3  assuming that only a 90\% of the transit signature is blocked by local \lya absorbers. Eq.~9 has been reworked for this purpose to,

\begin{equation}
3 = \left( \frac {\eta A_{eff} }{h \nu _{Ly\alpha}} \right) ^{1/2} \frac{\delta F^*_{Ly \alpha}}{F^*_{Ly \alpha}} 
\left( \frac { F^*_{Ly \alpha}} { 1+ 4F_{Back}/F^*_{Ly \alpha}}\right) ^{1/2}
\end{equation}
\noindent
being $F^*_{Ly \alpha}$ the stellar \lya flux, $\delta F^*_{Ly \alpha}$ the maximum variation of the \lya flux during the transit (the depth of the 
transit in the \lya light curve, see Figs.4-5) and $F_{back}$ the background \lya flux. This equation can be transformed into,

\begin{equation}
\eta  = \chi \left( \frac { 1+ 4F_{Back}/F^*_{Ly \alpha}}{ F^*_{Ly \alpha}} \right) 
\end{equation}
\noindent
with,
\begin{equation}
\chi = 9  \frac {h \nu _{Ly\alpha}}{ A_{eff} } \left( \frac{F^*_{Ly \alpha}}{\delta F^*_{Ly \alpha}}  \right) ^{2}
\end{equation}
\noindent
a constant that only depends on the depth of the transit, $\delta F^*_{Ly \alpha}/F^*_{Ly \alpha}$.
The curves for M5V and M0V stars in Fig.~8 are based on Proxima Centauri and AU~Mic light curves (see Fig.~5).
$T_0$ is the exposure time to reach SNR=3 for the Proxima Cen and AU~Mic \lya fluxes
(119~s for Proxima Cen and 4~h for AU~Mic). For instance, to detect the transit of an Earth-like planet
orbiting an M5 star with $F_{Ly\alpha} = 10^{-13}$~erg~cm$^{-2}$~s$^{-1}$ requires $\eta/\eta _0 = 40$ that
for instance, can be  achieved by increasing the size of the primary from 4 to 8 m and co-adding the signal of 10 transits.  
In a similar manner, it is required $\eta/\eta _0 = 100$ for an M0 star with $F_{Ly\alpha} = 10^{-13}$~erg~cm$^{-2}$~s$^{-1}$.

A quick search using the services of the  Centre de Donn\'ees Stellaires (Strasbourg, France) 
results in 80 stars within 5 pc around the Sun and 47 of them single M-type star. Ly$\alpha$ fluxes 
have been measured for few of them (Youngblood et al. 2016). There are two M5V stars in the sample,
GJ876 and GJ581, and their observed Ly$\alpha$ flux is $0.14\times 10^{-12}$~erg~s$^{-1}$~cm$^{-2}$
and  $0.04\times 10^{-12}$~erg~s$^{-1}$~cm$^{-2}$. According to  Fig.~7, the exosphere of an Earth-like
planet  would be easily detectable with an 8~m primary space telescope by the transit method.\par

 \begin{figure}[h!]
  \centering
  \includegraphics[width = .9\columnwidth]{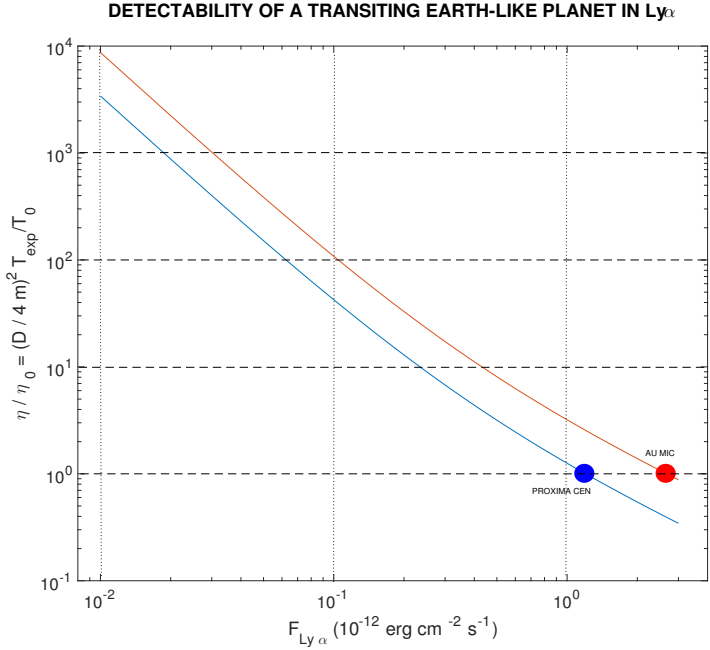}
  \caption{Detectability of a transiting Earth analogue by flux variations in the Ly$\alpha$ line as a function
  of the observed Ly$\alpha$ flux (Case B) and the diameter of the primary mirror of the telescope
  (see text for instrumentation configuration and detectability threshold definition).
}
  \label{fig:8}
\end{figure}

\section{Conclusions}
To summarise, it is feasible to measure the distribution of exospheric hydrogen in terrestrial
planets orbiting nearby cool M type stars with moderate size (4-8 m primary mirror)  space observatories. 
These measurements are required to understand the dominant processes in the upper
atmospheric layers and the energy flow between the atmosphere and the surrounding space 
because very different levels of space weather conditions can be tested.  In this manner, the role of
stellar activity and evolutionary state, i.e. the role of particle collisions and planetary magnetic 
fields in energising the neutral hydrogen atoms, could be addressed rigorously.



\end{document}